\documentclass{PoS}

\usepackage{latexsym}
\usepackage{amssymb}

\newcommand{\be}{\begin{equation}}
\newcommand{\ee}{\end{equation}}
\newcommand{\ba}{\begin{eqnarray}}
\newcommand{\ea}{\end{eqnarray}}
\newcommand{\bi}{\begin{itemize}}
\newcommand{\ei}{\end{itemize}}

\renewcommand{\>}{\rangle}

\newcommand{\fig}{Fig.~}

\newcommand{\la}{\label}

\newcommand{\Nf}{\mathop{N_{\rm f}}}

\newcommand{\txts}{\textstyle}

\newcommand{\dlangle}{\langle\!\langle}          
\newcommand{\drangle}{\rangle\!\rangle}          
\newcommand{\fslash}[1]{\slash\!\!\!{#1}}        
\newcommand{\ud}{\,\mathrm{d}}

\title{Moments of GPDs and transverse-momentum dependent PDFs from the lattice}

\ShortTitle{Moments of GPDs and transverse-momentum dependent PDFs from the lattice}

\author{\speaker{Harvey~B.~Meyer}%
\\
        Johannes Gutenberg Universit\"at Mainz, Institut f\"ur Kernphysik, D-55099 Mainz, Germany\\
        E-mail: \email{meyerh@kph.uni-mainz.de}}

\abstract{I review lattice-QCD calculations of 
the electromagnetic and generalized form factors (GFFs), which determine the 
transverse structure of the nucleon, and briefly comment on 
recent calculations related to 
transverse-momentum dependent parton distribution functions (TMDPDFs).}

\FullConference{XVIII International Workshop on Deep-Inelastic Scattering and Related Subjects\\
                 April 19-23, 2010\\
                 Convitto della Calza, Firenze, Italy}

\begin{document}

Lattice QCD calculations of hadron structure have been carried out for 
two decades, see~\cite{Hagler:2009ni} for a recent review. 
As the up- and down-quark masses are gradually lowered toward the physical
point 
(isospin symmetry is assumed),
new questions have arisen on the way.
The calculations have reached a certain degree of maturity and consensus 
among different collaborations down to about $m_\pi=300$MeV.
With the goal in mind of confronting the lattice results with hadron phenomenology,
it is useful to first discuss the quantities that are well determined 
experimentally.
Such quantities include the axial charge of the nucleon, 
the isovector momentum fraction and the electromagnetic (e.m.) form factors (FFs).
One may then focus on quantities where the lattice can be a discovery tool,
such as the generalized form factors.

%
%

I therefore start by discussing the calculation of e.m. FFs,
both because they are a special case of the GFFs
and because they represent a benchmark calculation of 
experimentally well-determined functions. The matrix elements of the 
e.m. current between two nucleon states are parametrized
by the Dirac and Pauli form factors ($q=P'-P$, $Q^2=-q^2$),
\ba
\<P',\Lambda'|J^\mu|P,\Lambda\> = \bar U(P',\Lambda') \Gamma^\mu(q^2)U(P,\Lambda)\,,~~~~~~~
\Gamma^\mu(q^2) = \gamma^\mu {F_1}(Q^2) + i\sigma^{\mu\nu}\frac{q_\nu}{2m}{F_2}(Q^2).
\ea
The result of a calculation at $m_\pi=300$MeV by the LHP collaboration
is displayed in \fig(\ref{fig:vmd}, top row).
It uses 2+1 flavors of domain-wall fermions at a lattice spacing of $a=0.084$fm.
Several other collaborations have obtained the FFs at similar pion 
masses~\cite{Yamazaki:2009zq}.
A common point among these calculations is that the dipole form fits the data
well, but the radii extracted from the fits are significantly smaller than the 
phenomenological radii. Secondly, chiral effective theory predicts a strong pion mass 
dependence of the Dirac and Pauli radii, but the lattice data exhibits a much 
milder dependence. It may be that the $m_\pi$-range of applicability of the chiral formulae
is much smaller than 300MeV. Finite-volume effects need to be investigated in more 
detail too.
%
%

The approximate dipole behavior of the phenomenological FF
$F_1(Q^2) = \frac{1}{1+Q^2/0.71{\rm GeV}^2}$ can be
understood as being due to the contribution of two nearby vector meson
poles with opposite residua~\cite{Perdrisat:2006hj}.
Figure (\ref{fig:vmd}, bottom row) illustrates the fact that in Nature, 
the Dirac dipole mass lies within $10\%$ of the $\rho$ meson mass.
In QCD at larger $m_{u,d}$ however, lattice results show that 
the Dirac dipole mass is significantly larger than the 
vector meson mass in the same theory.
%
We note that, since both the Dirac and 
Pauli radii diverge in the chiral limit, the latter must become
very large compared to the $\rho$ Compton wavelength approaching the chiral limit.

The numerical evaluation of Wick-connected and -disconnected diagrams
that contribute to three-point functions (\fig(\ref{fig:vmd}), bottom row) 
proceeds in a rather different way. Most collaborations have focused 
on the more tractable connected contributions, thus restricting 
themselves to isovector quantities.
Consider in this respect the flavor structure of the e.m. FFs of the 
proton\footnote{NB. The normalization is such that  $F_1^u(0)=2$, and
the Sachs FFs are given by $G_E=F_1-\frac{Q^2}{4M_N^2}F_2$, $G_M=F_1+F_2$.},
\ba 
F_{1,2} &=& {\txts\frac{2}{3}} F^u - {\txts\frac{1}{3}} F^d -{\txts\frac{1}{3}} F^s 
= {\txts\frac{1}{2}} F^{u-d} + {\txts\frac{1}{6}} F^{u+d-2s} 
= {\txts\frac{1}{2}} F^{u-d}_{\rm conn} +  {\txts\frac{1}{6}} F^{u+d}_{\rm conn}
+  {\txts\frac{1}{6}} (F^{u+d}_{\rm disc} -2F^s_{\rm disc}).
\nonumber
\ea
Figure (\ref{fig:vmd}, middle row), 
displaying the strangeness FFs~\cite{Doi:2009sq},
suggests that disconnected diagrams contribute less than $0.01$ 
to $F_1$, which is negligible at low $Q^2$ compared to the other uncertainties.

\begin{figure}
\vspace{-0.5cm}
\begin{minipage}{0.5\textwidth}
\centerline{\includegraphics[width=0.8\textwidth]{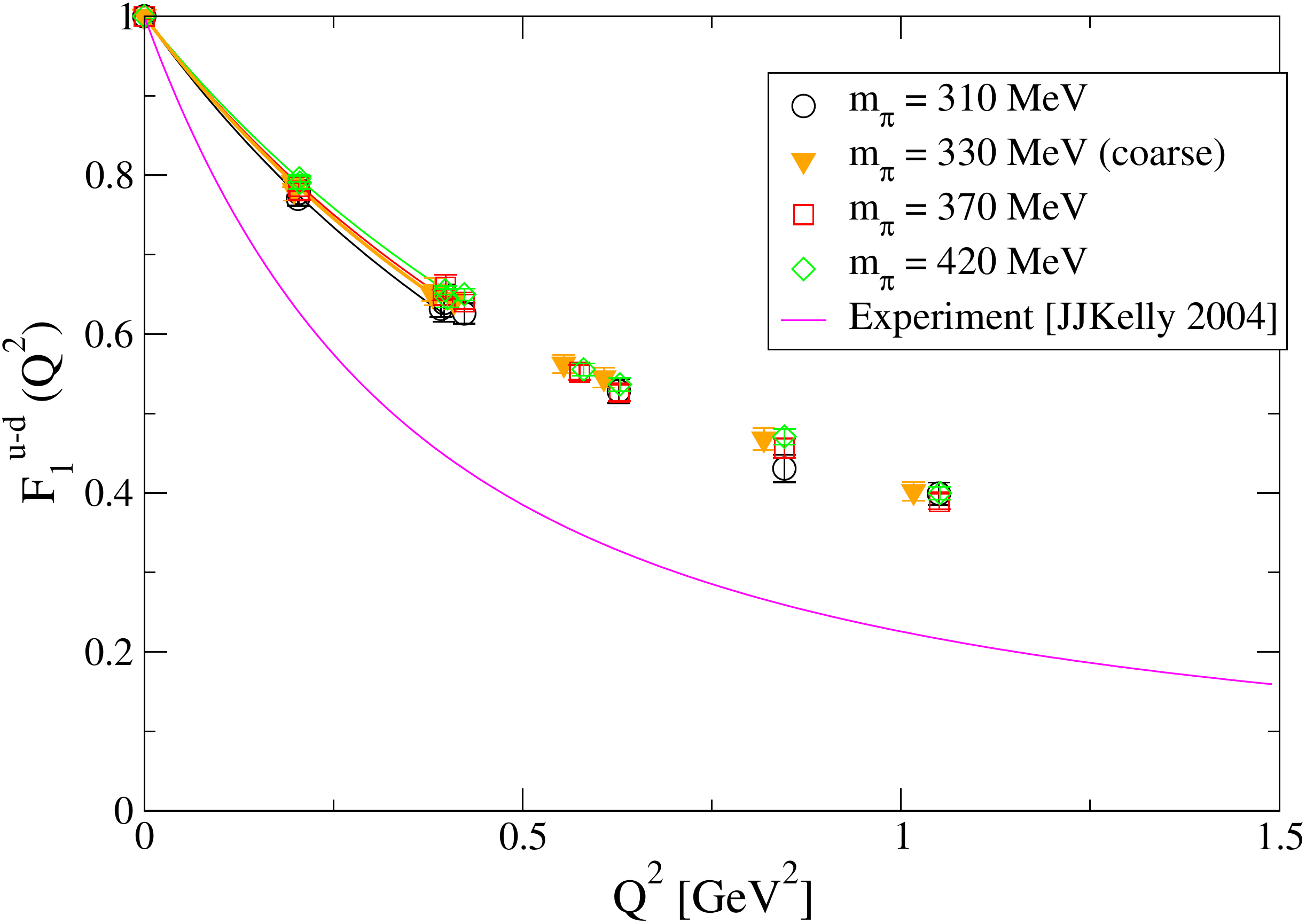}}
\end{minipage}
\begin{minipage}{0.5\textwidth}
\centerline{\includegraphics[width=0.8\textwidth]{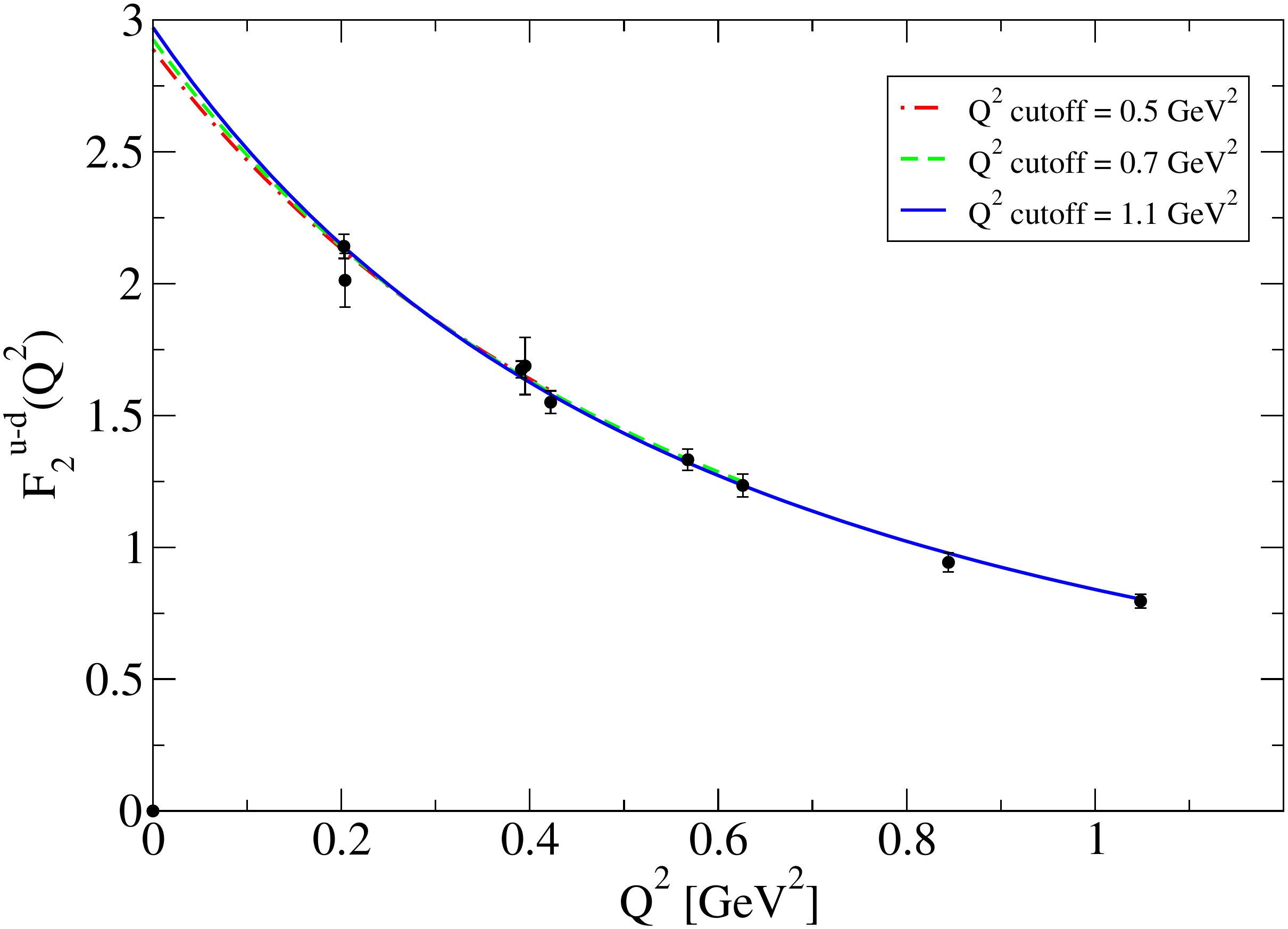}}
\end{minipage}
\begin{minipage}{0.5\textwidth}
\centerline{\includegraphics[width=0.68\textwidth,angle=-90]{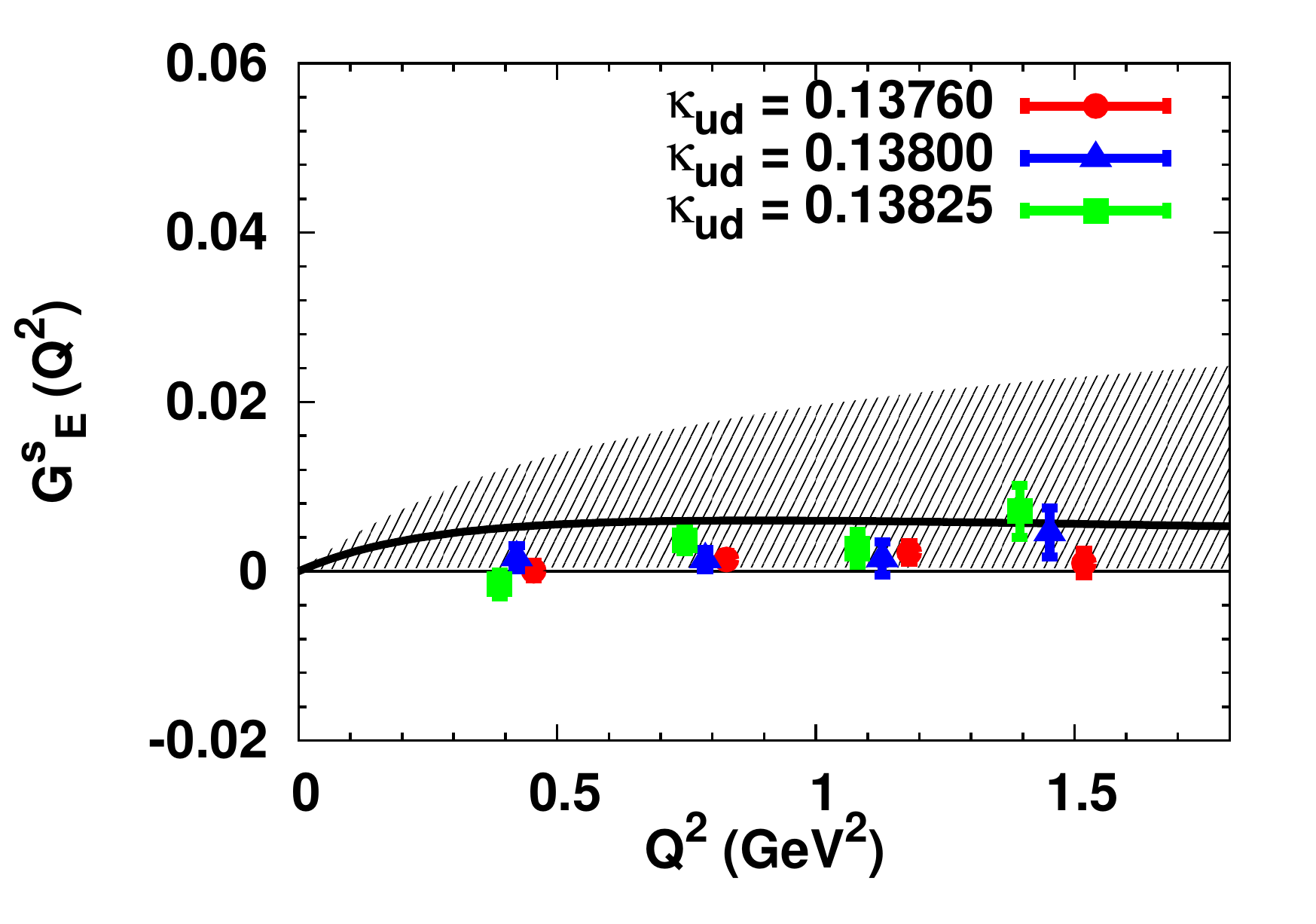}}
\end{minipage}
\begin{minipage}{0.5\textwidth}
\centerline{\includegraphics[width=0.68\textwidth,angle=-90]{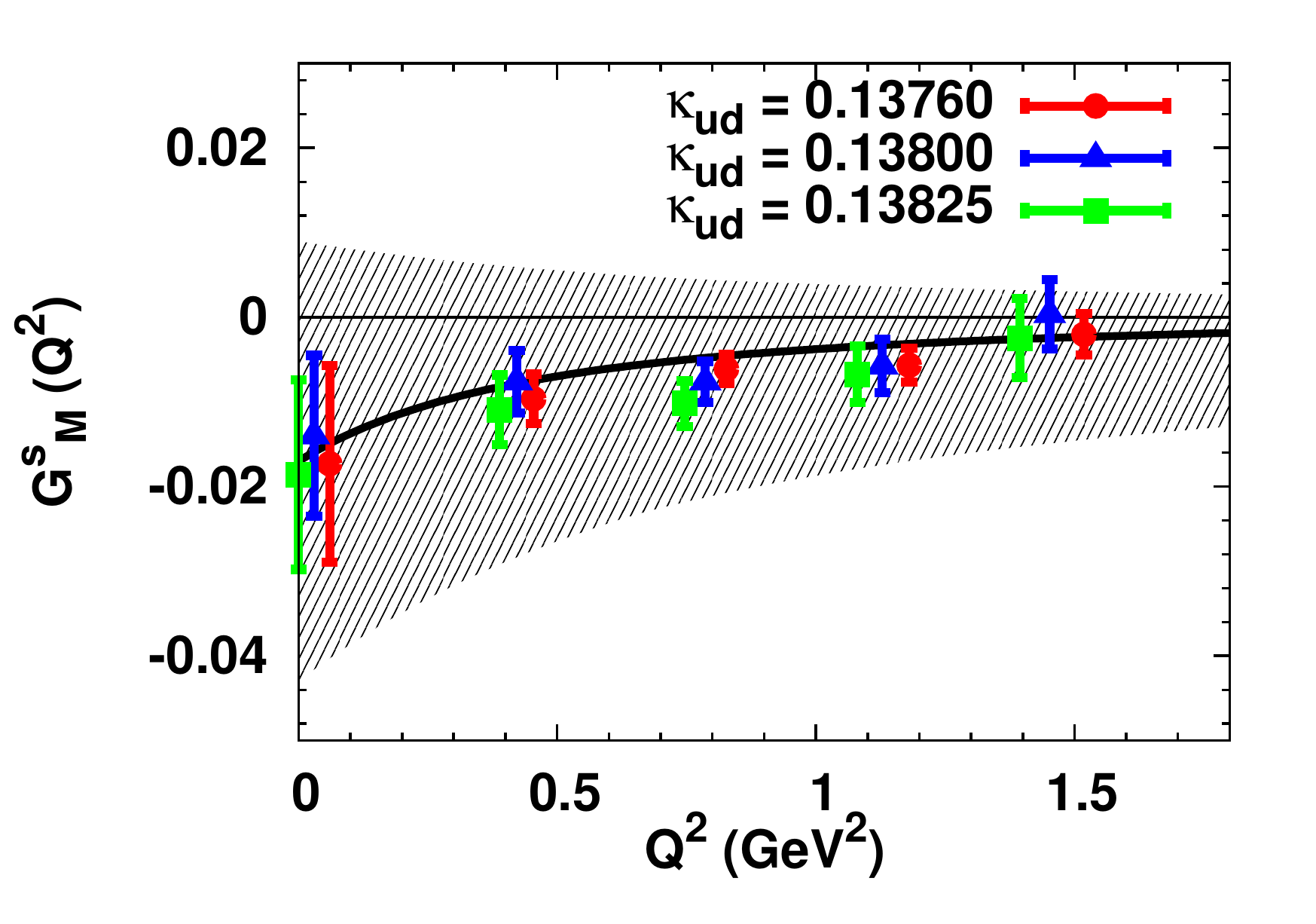}}
\end{minipage}
\vspace{-0.35cm}

\begin{minipage}{0.5\textwidth}
\centerline{\hspace{0.3cm}\includegraphics[width=0.5\textwidth]{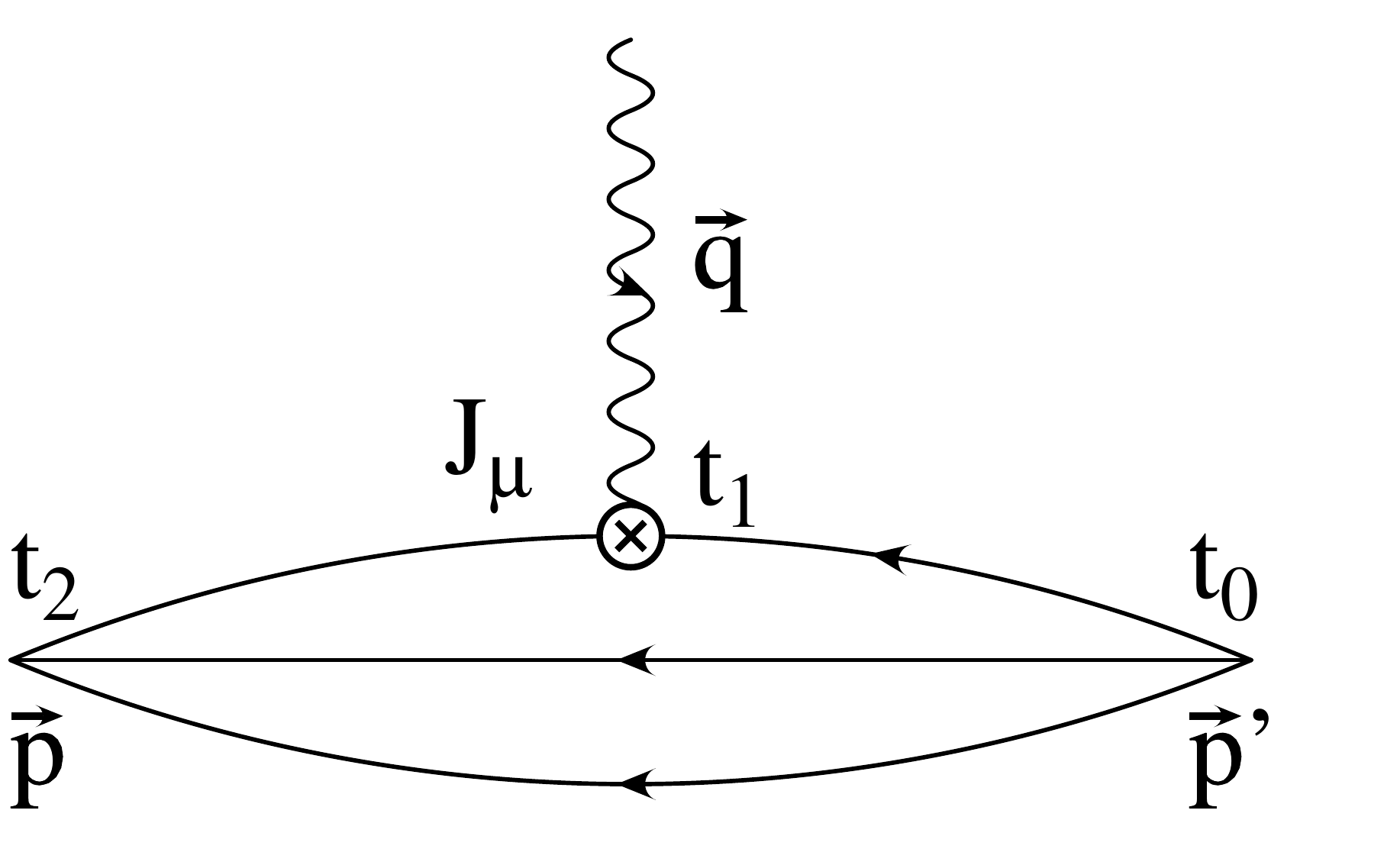}
            \includegraphics[width=0.5\textwidth]{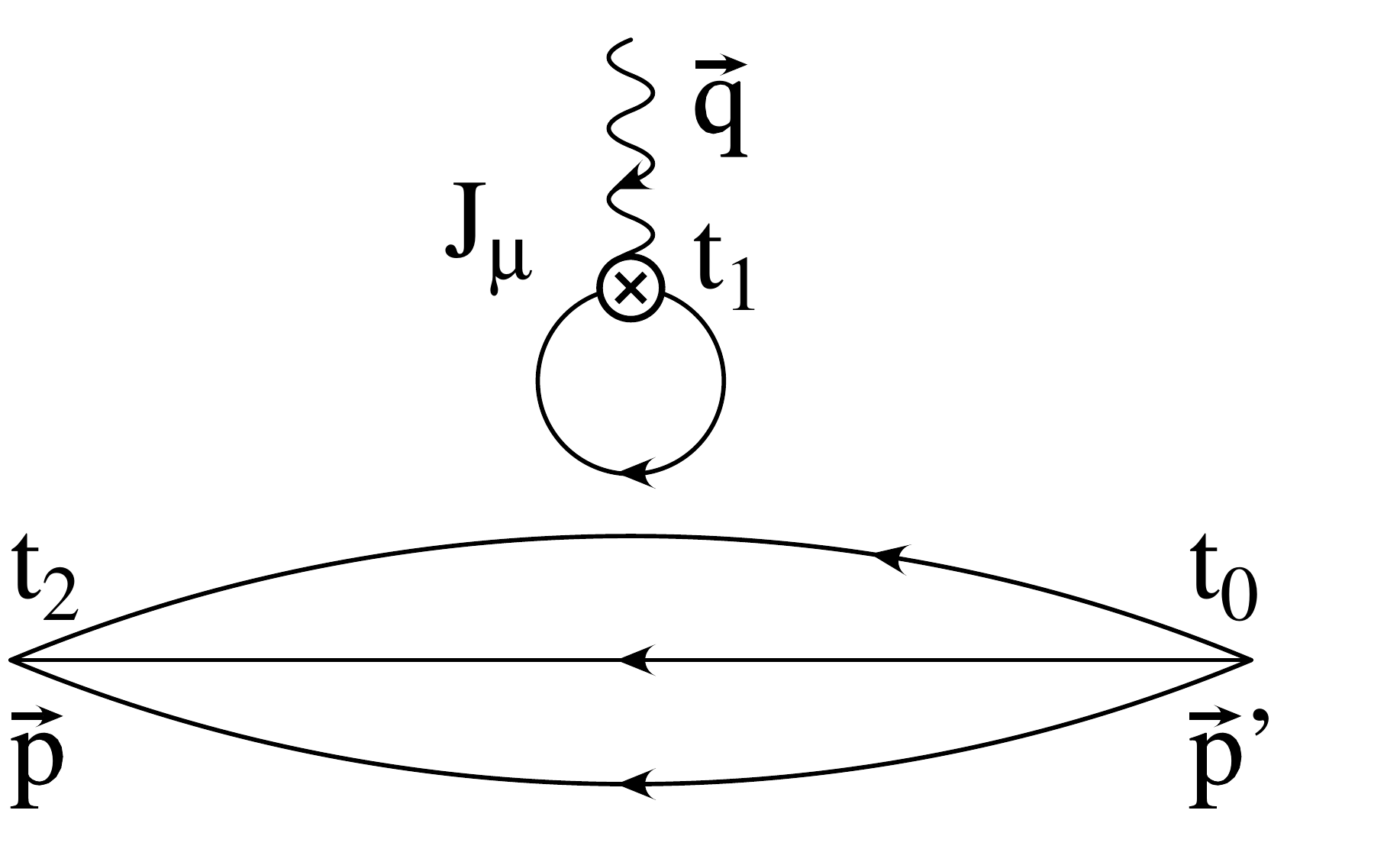}}
\end{minipage}
\begin{minipage}{0.5\textwidth}
\centerline{\hspace{-0.3cm}\includegraphics[width=0.97\textwidth]{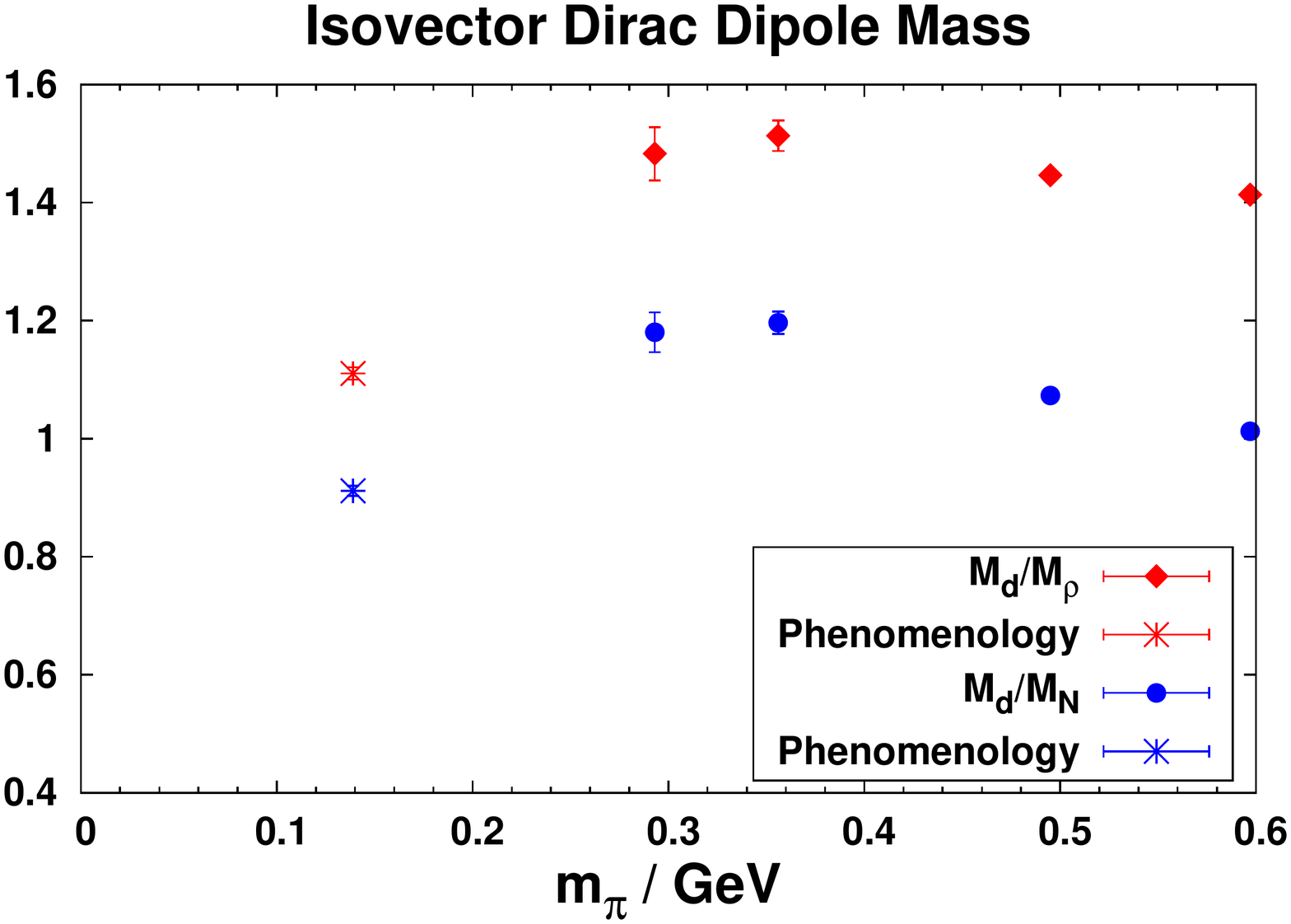}}
\end{minipage}
\vspace{-0.7cm}
\caption{
\underline{Top row}: the e.m.~FFs in an $\Nf=2+1$ calculation using 
domain-wall fermions at a lattice spacing 0.084fm (0.114fm in the `coarse' case).
The right plot displays the Pauli FF calculated 
at $m_\pi=300$MeV~\cite{Syritsyn:2009mx}.
\underline{Middle row}:  calculation of the strangeness FF 
of the nucleon by the $\chi$QCD collaboration~\cite{Doi:2009sq} 
($N_{\rm f}=2+1$, $a=0.12$fm, $m_\pi\geq 600$MeV).
\underline{Bottom row}: connected and disconnected Wick contractions contributing 
to hadronic matrix elements (fig. from~\cite{Doi:2009sq}).
Right, the nucleon Dirac dipole mass~\cite{Bratt:2010jn}.}
\la{fig:vmd}
\end{figure}
%
%
I now briefly describe the connection between the 
generalized parton distributions (GPDs), which govern 
Deeply Virtual Compton Scattering, and the GFFs that can be computed on the lattice.
The GPDs parametrize the matrix elements of the operators
\be
 {\cal O}_\Gamma(x) =\int \!\frac{d \lambda}{4 \pi} e^{i \lambda x} \overline
  q (\frac{-\lambda n}{2})
  \Gamma \,{\cal P} e^{-ig \int_{-\lambda / 2}^{\lambda / 2} d \alpha \, n
    \cdot A(\alpha n)}\!
  q(\frac{\lambda n}{2})\,.
\ee
Here $n$ is a light cone vector, $\fslash{n}\equiv n_\mu\gamma_\mu$,
and in the following $x$ is the momentum fraction, $\Delta\equiv P'-P$,
$t=\Delta^2$ and $\xi=-n\cdot\Delta/2$.
Different $\Gamma$ matrices lead to different GPDs,
\be
 \Gamma = \fslash{n}  \leadsto  (H,E),~~~~~
 \Gamma =\fslash{n} \gamma_5 \leadsto  (\widetilde H,\widetilde E),~~~~~
\Gamma = n_\mu \sigma^{\mu j}
       \leadsto  (H_T,E_T) ~{\rm and}~ (\widetilde H_T,\widetilde E_T).
\ee
For $\Gamma \doteq \fslash{n}$, the `unpolarized' case, the parametrization 
at a renormalization scale $\mu$ takes the form
\be
  \left\langle P',\Lambda ' \right|
  {\cal O}_{\not n}(x,\mu)
  \left| P,\Lambda \right \rangle
  = \dlangle\fslash{n}  \drangle H(x, \xi, t, \mu)
 + \frac{ n_\mu\Delta_\nu} {2 m}\dlangle i \sigma^{\mu \nu}\drangle E(x, \xi, t, \mu)\,,
\ee
where $\dlangle\Gamma\drangle\equiv \overline U(P',\Lambda ')\Gamma U(P,\Lambda) $
and $U(P,\Lambda) $ is a spinor that solves the free Dirac equation.
GPDs determine the distribution of partons and their
helicities in impact parameter space~\cite{Burkardt:2000za},
 \[
  q(x,{\bf b}_\perp) = 
 {\txts\int \frac{\ud^2{\bf b}_\perp}{(2\pi)^2}} \,
 e^{-i\Delta_\perp\cdot {\bf b}_\perp}\,
 H(x,\xi=0, -\Delta_\perp^2)\,,
 ~~~
  \Delta q(x,{\bf b}_\perp) = 
 {\txts\int \frac{\ud^2{\bf b}_\perp}{(2\pi)^2}} \,
 e^{-i\Delta_\perp\cdot {\bf b}_\perp}\,
 \widetilde H(x,\xi=0, -\Delta_\perp^2)\,.
 \]

Mellin-moments of the GPDs,
$ H^n(\xi, t) \equiv {\txts\int_{-1}^{1}} \ud x\, x^{n-1} H(x, \xi, t)$,
are given by polynomials in the longitudinal momentum transfer $\xi$,
\be
\begin{array}{l@{~~}l@{~~}l@{~~}l}
 H^{n=1}(\xi, t) &= {A_{10}(t)}, \quad& E^{n=1}(\xi, t) &= {B_{10}(t)},
\\
  H^{n=2}(\xi, t) &= {A_{20}(t)}+(2\xi)^2 {C_{20}(t)},\quad& E^{n=2}(\xi, t) &
= {B_{20}(t)}-(2\xi)^2 {C_{20}(t)},
\\
  H^{n=3}(\xi, t)  &= {A_{30}(t)}+(2\xi)^2 {A_{32}(t)}\,,\quad&
   E^{n=3}(\xi, t) &= {B_{30}(t)}+(2\xi)^2 {B_{32}(t)}\,,~\dots
\end{array}
\ee
The coefficients of these polynomials are the GFFs
calculated on the lattice. In the $n=1$ case, $A_{10}$ and $B_{10}$ 
coincide with $F_1$ and $F_2$. In the $n=2$ unpolarized case, the operator is
${\cal O}_{\mu_1\mu_2}(x) = \frac{1}{2}\bar q 
(\gamma_{\mu_1} {D}_{\mu_2}  
 + \gamma_{\mu_2} {D}_{\mu_1}) 
q \,$ and, with $\bar P \equiv (P+P')/2 $,
\be
\langle P'| {\cal O}_{\mu_1\mu_2}(q) | P \rangle =
\bar P_{\lbrace\mu_1}\dlangle  
\gamma_{\mu_2\rbrace}\drangle\,{A_{20}(t)}
+ {\txts\frac{i}{2 m}} \bar P_{\lbrace\mu_1} \dlangle
  \sigma_{\mu_2\rbrace\alpha}\drangle \Delta_{\alpha}\, {B_{20}(t)}
 +{\txts\frac{1}{m}} 
\Delta_{\{ \mu_1}   \Delta_{ \mu_2 \} }\,  {C_{20}(t)}\,.
\ee
The forward-limit of $A_{20}$ and $B_{20}$ determine the 
momentum and spin decomposition of the nucleon in terms of 
quark and gluon contributions~\cite{Ji:1996ek}.
In practice, only the first few moments $n\leq 3$ 
are accessible on the lattice. 
The impact-parameter dependence ${\bf b}_\perp$ can be 
obtained more easily, 
currently $ (0.4{\rm GeV})^2<\Delta_\perp^2<(2.0{\rm GeV})^2$.
The lower end of this interval is 
limited by momentum quantization in a finite box,
$P=\frac{2\pi k}{L}$, $k\in \mathbb{Z}$.
It can potentially be overcome with twisted boundary conditions, 
whereby $P= \frac{\theta}{L}+\frac{2\pi k}{L}$~\cite{Gockeler:2008zz}.
The upper end is limited by the inverse lattice spacing.

Figure (\ref{fig:transv-r}, left panel) 
displays the $n=1,2$ and 3 GFFs of the nucleon (isovector, unpolarized case). 
This particular calculation is performed in a (3.5fm)$^3$ box at a pion mass of 355MeV.
For the purpose of the comparison, 
the FFs have been normalized by their forward value. 
It is clearly seen that the FFs become flatter as $n$ increases. 
This translates into decreasing transverse radii,
(right panel). The latter have been obtained by performing 
dipole fits to the FF data with $|t|\leq0.5{\rm GeV}^2$.
The simulations are performed in boxes of size (2.5fm)$^3$, 
with in addition a (3.5fm)$^3$ simulation at $m_\pi=355$MeV.
In the $n=1$ case, a gradual increase in the (Dirac) radius 
is observed as the pion mass is lowered toward its physical value.
For $n\geq2$, this growth in the transverse size of the nucleon is 
not observed in the range of explored pion masses.
This is qualitatively consistent with the expectation 
from chiral effective theory,
which predicts a logarithmic divergence in the $n=1$ case, but 
finite values for the radii corresponding to higher $n$.
However, at $m_\pi=355$MeV 
one also observes a statistically significant difference between
the radii extracted from volumes (2.5fm)$^3$ and (3.5fm)$^3$.
Since the finite-size effect is expected to increase when the pion mass
decreases, it appears necessary to check the observed trends by using
large-volume simulations.


\begin{figure}
\vspace{-0.7cm}
\begin{minipage}{0.5\textwidth}
\centerline{\includegraphics[width=0.85\textwidth]{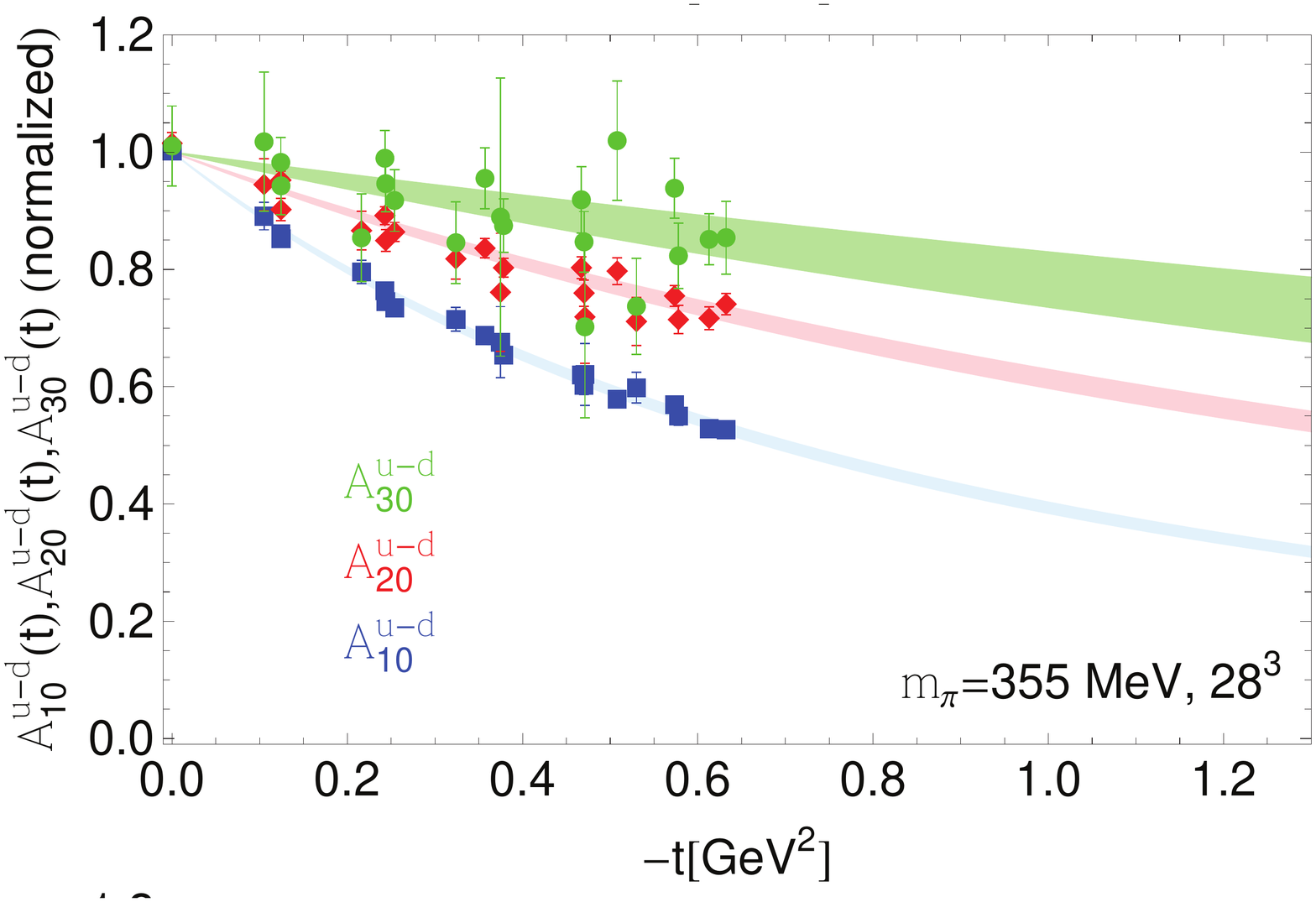}}
\end{minipage}
\begin{minipage}{0.5\textwidth}
\centerline{\includegraphics[width=0.85\textwidth]{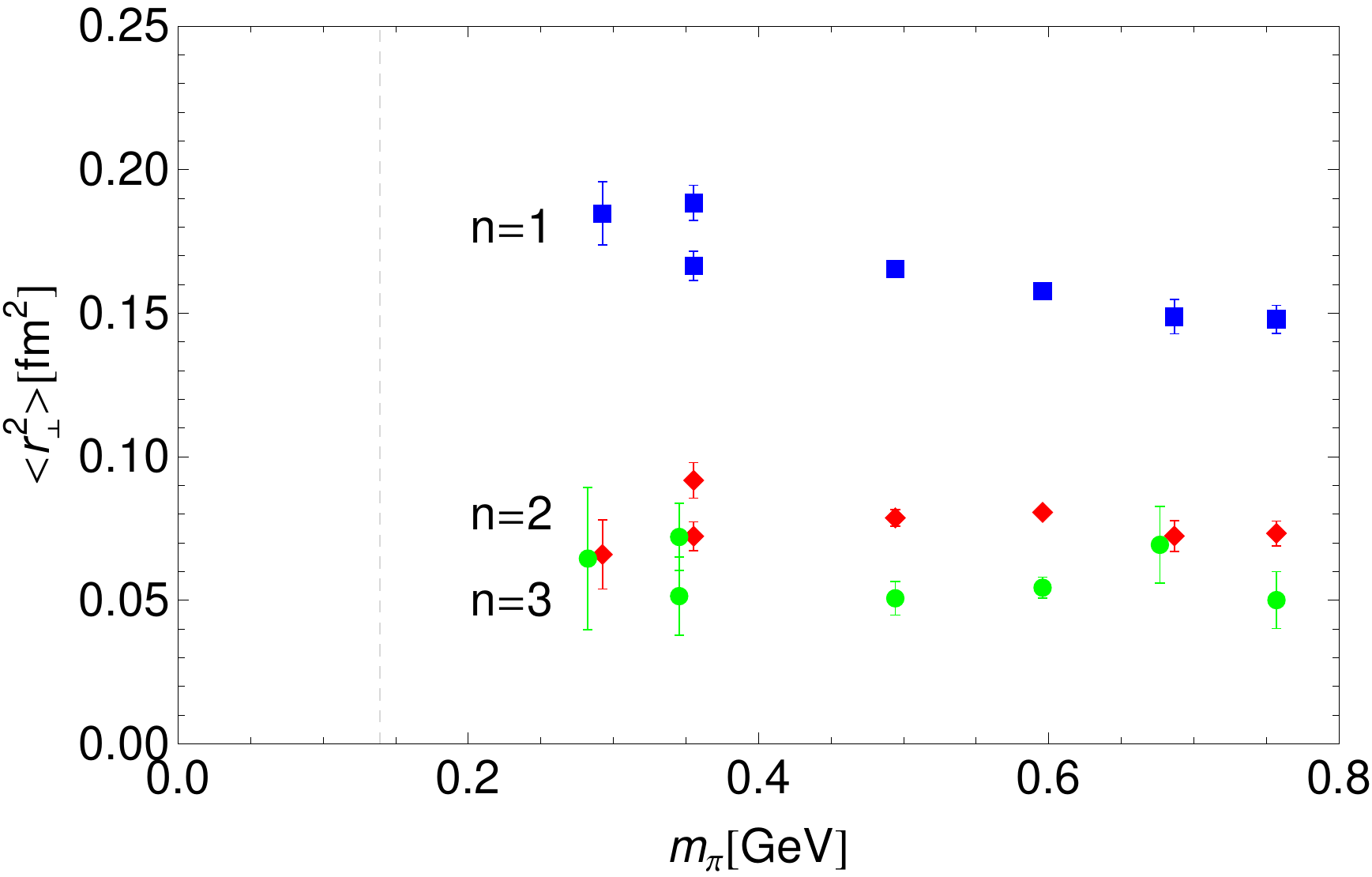}}
\end{minipage}
\vspace{-0.5cm}
\caption{Left: Unpolarized isovector GFFs.
         Right: the corresponding radii~\cite{Bratt:2010jn}.}
\la{fig:transv-r}
\end{figure}


In conclusion, GPDs play a central role in describing quantitatively 
the three-dimensional structure of the nucleon.
Lattice QCD is ideally suited to determine the GFFs
associated with their low Mellin-moments.
A handful of lattice collaborations have published 
results for the nucleon structure for pion masses down to 
about 300MeV. A general observation is that 
many observables show a very mild pion mass dependence.
The non-analytic pion-mass dependence 
predicted by chiral effective theory is not yet
visible in the data; presumably it
sets in at a lower pion mass.
\nopagebreak
The next step will therefore be to go down to 200MeV~\cite{Gockeler:2009pe}
with controlled uncertainties, in particular a solid understanding 
of the finite-volume effects in the simulations is crucial.

The formalism to study transverse-momentum dependent PDFs 
on the lattice, relevant to semi-inclusive DIS, has recently 
been studied~\cite{Hagler:2009mb}. One important aspect 
of the calculation is the renormalization of non-local operators 
of the type $\bar q(x)\Gamma\,U_{x,0}q(0)$, where $U_{x,0}$ 
is a Wilson line. A second essential issue is the connection between
the matrix elements of this operator with $U_{x,0}$
along the light-cone and with $U$ in the Euclidean domain.
I expect to see further developments in this field in the near future.


\bibliographystyle{JHEP}
\bibliography{/home/meyerh/CTPHOPPER/ctphopper-home/BIBLIO/viscobib.bib}


\end{document}